\documentclass[11pt,a4paper]{article}
\usepackage[a4paper]{geometry}
\usepackage{graphicx}
\usepackage{amsmath,amssymb,amsthm,mathtools}
\usepackage{enumitem}
\usepackage{xcolor}
\usepackage[
    colorlinks=true,
    linkcolor=blue,
    citecolor=blue,
    urlcolor=blue
]{hyperref}
\usepackage[nameinlink,noabbrev]{cleveref}
\usepackage{microtype}
\usepackage{xcolor}

\newcommand{\Fact}{\mathrm{Fact}}
\newcommand{\rank}{\mathrm{rank}}
\newcommand{\pc}{\mathrm{pc}}
\newcommand{\eps}{\varepsilon}

\theoremstyle{plain}

\newtheorem{theorem}{Theorem}[section]
\newtheorem{lemma}[theorem]{Lemma}

\theoremstyle{definition}

\newtheorem{definition}[theorem]{Definition}

\theoremstyle{remark}

\usepackage{fixme}
\fxsetup{
    status=final,
    author=,
    layout=footnote, 
    theme=color
}

\title{The Regular Cover Problem is NP-complete}

\author{
Stefan Hetzl\footnote{corresponding author}\\
\url{stefan.hetzl@tuwien.ac.at}\\
Institute of Discrete Mathematics and Geometry\\
TU Wien, Austria
\and
Parham Shadman\\
\url{parhammshadman@gmail.com}\\
Institute of Discrete Mathematics and Geometry\\
TU Wien, Austria
}

\begin{document}

\maketitle

\begin{abstract}
The regular cover problem is the following question: given a finite language $L$ and
 a $k\geq 0$, is there an acyclic regular grammar $G$ with at most $k$ productions
 that covers $L$, i.e., that satisfies $L(G) \supseteq L$.
The study of this problem is motivated by applications in proof theory and automated
 deduction.
The regular cover problem is known to be in NP.
In this paper we show that it is NP-complete by reducing the vertex cover problem
 for graphs with maximum degree three to it.
\end{abstract}

\section{Introduction}

The grammatical complexity of a formal language is the complexity
 of a minimal grammar generating this language.
The study of the grammatical complexity of formal languages was started in~\cite{Gruska67Classification}.
This line of research has been continued 
 in~\cite{Gruska69Some,Gruska71Complexity,Gruska72Size,Cerny77Complexity}, where, among 
 others, the number of productions of a grammar has been considered as complexity measure.
In~\cite{Bucher81Concise}, a theory of the grammatical complexity of finite languages in
 terms of production complexity was initiated by giving a relative succinctness classification for various kinds of context-free grammars.
Investigations along these lines have been continued in, e.g.,~\cite{Alspach83Lower,Bucher81Note,Bucher84Context,Tuza87Context,Dassow12Production,Dassow17Descriptional}.

The cover complexity of a finite language $L$ is the minimal number of productions of a 
 grammar $G$ such that $L(G)$ is finite and $L(G) \supseteq L$.
Note that this condition is similar to (but different from) the one imposed on cover
 automata~\cite{Campeanu98Minimal,Campeanu01Minimal}.
The cover complexity of finite languages has been studied 
 in~\cite{Hetzl18Cover,Hetzl19Cover,Wolfsteiner20Grammatical}.
Our interest in this notion is primarily motivated by applications in proof theory.
As shown in~\cite{Hetzl12Applying}, there is an intimate relationship between a
 certain class of formal proofs (those with $\Pi_1$ cuts) in first-order predicate logic
 and a certain class of grammars (totally rigid acyclic tree grammars (TRATG)). 
This relationship has been exploited for a number of results in proof theory and automated
 deduction~\cite{Hetzl12Towards,Hetzl14Algorithmic,Hetzl14Introducing,Ebner19Generation}.
The central automata-theoretic problem on which most of this work is based is the 
 computation of a grammar that covers a given language with a minimal number of productions.
A number of practical algorithms have been developed, see, 
 e.g.,~\cite{Eberhard17Algorithmic} and a concrete example for a cover-incompressible 
 language has been given in~\cite{Eberhard15Compressibility,Eberhard18Compressibility}.
Phrased as decision problem this becomes the {\em regular cover problem}: given a finite
 language $L$ and a $k\geq 0$, is there an acyclic regular grammar with at most $k$ productions s.t.\ $L(G) \supseteq L$.
The complexity of this and many related decision problems has been studied
 in~\cite{Eberhard2018} where it was shown that the regular cover problem is in NP
 but it could only be conjectured that it is also NP-complete.

In this paper we prove that the regular cover problem is NP-complete.
Our proof proceeds by a reduction of the vertex cover problem on graphs of maximum
 degree at most three.
As stated in \cite{Eberhard2018}, this result implies the other conjecture in 
 \cite{Eberhard2018}, namely, the NP-completeness of the TRATG cover problem.
Moreover, we prove that the regular cover is NP-complete even for a fixed alphabet
 of size $2$.
Our work also uses some notions from~\cite{Neraud1990}\fxnote{There should be a thorough
 discussion of the relationship between our work and that of Neraud and papers citing it.}.


\section{Preliminaries}

For a word $w_1 , w \in \Sigma^*$, we say $w_1$ is a factor of $w$
 if there exists $u , v \in \Sigma^*$ such that $w = u w_1 v$.
For a finite set $L$ of words, define $\text{Fact}(L) =
 \{\text{nonempty factors of words in } L\}.$
Let $Y \subseteq \Sigma^*$\fxnote{why not $Y \subseteq \Sigma^+$}
 and $w \in Y^*$, a factorization of $w$
 over $Y$ is a tuple $(y_1, \dots, y_s)$ with $y_i \in Y \setminus \{\eps\}$ and $w = y_1 \cdots y_s$.
The elements $y_i$ are called {\em tokens}.
For a finite $L \subseteq \Sigma^+$, define $\rank(L) = \min\{|Y| \mid
 Y \subseteq \Sigma^+, \ L \subseteq Y^*\}$\fxnote{say where this notion comes from, cite~\cite{Neraud1990}?}.

\begin{definition}[Acyclic Regular Grammar]
A {\em regular grammar $G = (A, N, \Sigma, P)$} is a tuple consisting of a start symbol
 $A \in N$, a finite set of nonterminals $N$, a finite set of letters $\Sigma$ such that
 $N \cap \Sigma = \emptyset$, and a finite set of productions
 $P \subseteq N \times \Sigma^*(N \cup \{\eps\})$.
We call $G$ {\em acyclic} if there exists a strict linear order $\prec$ on the nonterminals
 such that $B \prec C$ whenever $B \rightarrow wC \in P$ for some $w \in \Sigma^*$.
\end{definition}

\begin{definition}[Regular Cover Problem]
Given a finite set of words $L$ and $k \geq 0$, is there an acyclic regular grammar $G$ such
 that $|P(G)| \leq k$ and $L \subseteq L(G)$?
\end{definition}

\begin{definition}
For a finite simple graph $H = (V, E)$, a set $C \subseteq V$ is a vertex cover if every
 edge has at least one endpoint in $C$.
The minimum possible size of a vertex cover is denoted by $\tau(H)$.
\end{definition}

\section{NP-Completeness of the Regular Cover Problem}

\subsection{Grammar construction}

We begin by proving the following lemma, which gives a lower bound on the number of productions of an acyclic regular grammar $G$ covering a finite language $L$ based on $\rank(L)$

\begin{lemma}\label{lem.rank_numprods}
Let $L$ be a finite set of nonempty words over $\Sigma$ and let $G = (A, N, \Sigma, P)$ be
 an acyclic regular grammar with $L \subseteq L(G)$.
Then
\[
|P| \geq \rank(L).
\]
\end{lemma}
\begin{proof}
For every word $w \in L $, choose a derivation of $ w $ in $ G$.
Let $P_w \subseteq P$ be the set of productions occurring in this derivation.
Define $F_w= \{w' \in \Sigma^* \mid N_1 \to w' \in P_w  \text{ and }  N_1 \in N\} \cup \{w' \in \Sigma^* \mid N_1 \to w' N_2 \in P_w \text{ and } N_1,N_2 \in N\}$
 and let $F = \bigcup_{w \in L} F_w$.
Then every $w \in L$ is a concatenation of some elements of $F$, hence $L \subseteq F^*$.
And since every element of $F$ comes from at least one production, we have 
\[
\rank(L) \leq |F| \leq |P|.
\]
\end{proof}

For a finite language $L \subseteq \Sigma^*$ and a set $Y \subseteq \Sigma^*$ such that $L \subseteq Y^*$, our aim is to construct an acyclic regular grammar $G$ from $Y$ such that $L \subseteq L(G)$, and to derive an upper bound on the number of productions required to generate $L$ in terms of the size of $Y$. With this goal in mind, we present the following definition:
\vspace{5pt}

\begin{definition}
    A level assignment for an alphabet $\Sigma$ is a map
    \[
    \lambda : \Sigma \to \mathbb{Z} \times \mathbb{Z}, \qquad \lambda(a) = (\mathrm{in}(a), \mathrm{out}(a)),
    \]
    with
    \[
    \mathrm{in}(a) < \mathrm{out}(a).
    \]
    A word $a_1 \cdots a_n$ is called {\em $\lambda$-chained} if $\mathrm{out}(a_i) = \mathrm{in}(a_{i+1}) \text{ for all } i < n$. A finite language is called a {\em $\lambda$-chained} if all of its words are $\lambda$-chained. and also define:
    \[
    I(\lambda) = \{\mathrm{in}(w) : w \in L\}, \qquad O(\lambda) = \{\mathrm{out}(w) : w \in L\}.
    \]
    we call every integer in $I \cup O$ a {\em level}.
\end{definition}
\vspace{5pt}

\begin{lemma}\label{lem.lchained_grammar}
    Suppose $L$ is $\lambda$-chained, and
    \[
    L \subseteq Y^*.
    \]
    Then one can construct an acyclic regular grammar with at most
    \[
    |Y| + |I(\lambda)| + |O(\lambda)|
    \]
    productions.
\end{lemma}

\begin{proof}
    for every relevant level $l$ introduce a nonterminal $N_l$, and add the following productions to G:
    \[
    N_{\mathrm{in}(a_1)} \to w N_{\mathrm{out}(a_n)} \qquad (w=a_1 \cdots a_n \in Y),
    \]
    \[
    A \to N_h \qquad (h \in I),
    \]
    \[
    N_h \to \eps \qquad (h \in O).
    \]
    $G$ is acyclic because every derivation moves from a level to a higher level. It can be easily seen that $L \subseteq L(G)$ and $|P(G)| = |Y| + |I| + |O|$.
\end{proof}

\subsection{Reduction}\label{sec.reduction_unbalph}

For\fxnote{The things explained in this paragraph should be done more formally.} a graph $H = (V, E)$ where $V = \{v_1, \dots, v_n\}$ and the degree of each vertex is at most 3, we will introduce a language $L_H$ as follows:
Consider an arbitrary proper 4-coloring of the vertices, which can be computed in polynomial time by a greedy algorithm, represented by a mapping $C: V \to \{1, 2, 3, 4\}$. For every edge $e \in E$ consider the private letters $a_e, m_e, b_e$ and let
\[
\Sigma = V \cup \bigcup_{e \in E} \{a_e, m_e, b_e\}
\]
and for every edge $e = (u, v)$ with $C(u) < C(v)$ consider the set
\[
S_e = \{u m_e v, u m_e b_e, a_e m_e v, a_e m_e b_e\}
\]
and let
\[
L_H = \bigcup_{e \in E} S_e
\]

\begin{lemma}\label{lem.LH_lchained}
    $L_H$ is $\lambda$-chained.
\end{lemma}

\begin{proof}
    Let $ C$ be the fixed coloring computed in polynomial time as stated above.
    For $u \in V$ define $\lambda(u) = (2C(u)-1 \,,\,  2C(u))$. And for private letters of $e=(u,v)$ where $C(u) < C(v)$ let 
    \[
    \lambda(a_e) = \lambda(u), \qquad \lambda(b_e) = \lambda(v),
    \]
    and
    \[
    \lambda(m_e) = (2C(u), 2C(v) - 1).
    \]
    Every word in $L_H$ is $\lambda$-chained and hence $L_H$ is a $\lambda$-chained language. 
\end{proof}
\vspace{5pt}

Each private letter is related to exactly one edge. Any factor of a word in $L_H$ containing a private letter from edge $e$ cannot also contain a private letter belonging to a different edge. For a token set $Y \subseteq \Fact(L_H)$ and an edge $e$, define
\[
Y_e = \{y \in Y : y \text{ contains a letter from } \{m_e, a_e, b_e\}\}.
\]
Then
\[
Y = (Y \cap V) \mathbin{\dot{\cup}} \bigcup_{e \in E} Y_e.
\]

\begin{lemma}\label{lem.Y_and_Se}
Let $Y \subseteq \Fact (L_H)$ and $e=(u,v)$.
If $Y^*$ covers $S_e$, i.e., $S_e \subseteq Y^*$, then:
\begin{enumerate}
    \item \textit{$|Y_e| \geq 3$.}
    \item \textit{If neither $u$ nor $v$ belongs to $Y$, then $|Y_e| \geq 4$.}
    \item \textit{If $u \in Y$, then the three private tokens}
    \[
    \{a_e, m_e v, m_e b_e\}
    \]
    \textit{together with $u$ cover $S_e$. If $v \in Y$, then}
    \[
    \{u m_e, a_e m_e, b_e\}
    \]
    \textit{together with $v$ cover $S_e$.}
\end{enumerate}
\end{lemma}

\begin{proof}
Consider the three claims.
    \begin{enumerate}
    
    \item In $u m_e b_e$, consider a token covering the final $b_e$. It must contain $b_e$ and therefore cannot contain $a_e$.
    In $a_e m_e v$, consider a token covering the initial $a_e$. It contains $a_e$ and therefore cannot contain $b_e$.
    In $u m_e v$, consider a token covering the middle $m_e$. It contains neither $a_e$ nor $b_e$.
    Therefore, these three tokens are pairwise distinct, and all belong to $Y_e$. Hence
    \[
    |Y_e| \geq 3.
    \]
    
    \item Assume $u, v \notin Y$
    Look at $u m_e v$. Its first token cannot be the singleton $u$. Hence the first token is either $u m_e$ or the whole word $u m_e v$. If the first token were $u m_e $, the remaining suffix $v$ would have to be a token, contradicting $v \notin Y$. Thus $ u m_e v \in Y$
    Now inspect the other two words. The word $u m_e b_e$ forces either the whole word $u m_e b_e$ to be a token or the pair $u m_e, b_e$ to occur. Similarly, $a_e m_e v$ forces either the whole word $a_e m_e v$ or $a_e$ and $m_e v$.
    
    All these forced tokens are distinct from $u m_e v$. Finally, $a_e m_e b_e$ must also be covered, and any tokenization of it yields at least one more distinct token. Therefore $|Y_e| \geq 4$.
    
    \item If $u \in Y$, use
    \begin{align*}
    u m_e v &= u \cdot m_e v, \\
    u m_e b_e &= u \cdot m_e b_e, \\
    a_e m_e v &= a_e \cdot m_e v, \\
    a_e m_e b_e &= a_e \cdot m_e b_e.
    \end{align*}
    and the other case is similar.
    \end{enumerate}
\end{proof}

\begin{lemma}\label{lem.rank_LH}
    For a graph $H= (V, E)$ of maximum degree 3, $\rank(L_H)=3|E|+\tau(H)$.
\end{lemma}
\begin{proof}
    For a $Y \subseteq \Sigma^*$ such that $L_H \subseteq Y^*$, let $S = Y \cap V$ and let $U(S)$ be the set of edges that are not covered by $S$ in $V$. By Lemma~\ref{lem.Y_and_Se} for every $e \in E$ covered by $S$ we have $|Y_e| \geq 3$, and for every uncovered $e \in E$ we have $|Y_e| \geq 4$. therefore
    \[
    |Y| = |S| + \sum_{e \in E} |Y_e| \geq |S| + 3|E| + |U(S)|.
    \]
    and since $|S| + |U(S)| \geq \tau(H)$ for every $S \subseteq V$. then 
    \[
    |Y| \geq 3|E| + \tau(H).
    \]
    Conversely, let $C \subseteq V$ be a minimum vertex cover, for each edge choose the three private tokens introduced in Lemma~\ref{lem.Y_and_Se}(3). Their union with $C$ covers $L_H$ and its size is $|C| + 3|E| = \tau(H) + 3|E|$.
\end{proof}

\begin{theorem}\label{thm.regcover_npcomplete}
The Regular Cover Problem is NP-Complete.
\end{theorem}

\begin{proof}
As shown in~\cite{Eberhard2018}, the Regular Cover problem is in NP.
We prove the NP-hardness.

Let $(H,k)$ be an input for the Vertex Cover Problem where $H=(V, E)$ is a graph with n vertices 
 of maximum degree 3 and $m$ edges.
Let $H'=(V', E')$ be the disjoint union of $7$ copies of $H$ which can be constructed in 
 polynomial time from $H$.
Then it is easy to see that $H$ has a vertex cover of size less than or equal to $k$ if and
 only if $H'$ has a vertex cover of size less than or equal to $7k$. 
 
Now construct $L= L_{H'}$ which also can be constructed in polynomial and choose $\lambda$ as 
 introduced in Lemma~\ref{lem.LH_lchained} which also can be computed in polynomial time and
 for $L$ and $\lambda$ every word in $L=L_{H'}$ the possible starting levels belong to
 $\{1, 3, 5\}$ and possible ending levels belong to $\{4, 6, 8\}$ and hence
 $|I(\lambda)|+|O(\lambda)| \leq 6$.
Let $N = 3(7)m + 7k + 6$.
We will prove that $H$ has a vertex cover of size less than or equal to $k$ if and only if
 there exists an acyclic regular grammar $G$ such that $|P(G)| \leq N$ and $L \subseteq L(G)$.
 
For the left-to-right direction, suppose $H$ has a vertex cover of size $k$ then $H'$ has a
 vertex cover of size $7k$ and by Lemma~\ref{lem.rank_LH}
\[
\rank(L)=21m+7\tau(H) \leq 21m+7k.
\]
    
Suppose that $F$ is a minimum token set such that $L \subseteq F^*$.
Then $|F| = \rank(L)$.
By Lemma~\ref{lem.lchained_grammar} there exists an acyclic regular grammar G such that
 $L \subseteq L(G)$ with
\[
|P(G)| \leq |F|+|I(\lambda)|+|O(\lambda)| \leq \rank(L)+6 \leq 21m+7k+6 = N.
\]

For the right-to-Left direction, suppose that there exists an acyclic regular grammar G\fxnote{$N$ is used for nonterminals and for a natural number. This should be avoided.}
 with $|P(G)| \leq N$ and $L \subseteq L(G)$. By Lemma~\ref{lem.rank_numprods}:
    \[
    \rank(L) \leq |P(G)|
    \]
    \[
    21m+7\tau(H) \leq 21m+7k+6
    \]
    \[
    7\tau(H)\leq 7k+6
    \]
    \[
    \tau(H) \leq k+ \frac{6}{7} < k+1
    \]
    and since $\tau(H)$ is an integer, we have $\tau(H) \leq k$.
\end{proof}

\section{NP-Completeness of the Regular Cover Problem for Fixed Alphabet}

In this section, we will encode the language $L_H$ from the proof of 
 Theorem~\ref{thm.regcover_npcomplete} as a language over the alphabet $\{0, 1\}$
 and the factors are allowed to have overlaps.

\begin{definition}
For a finite language $L\subseteq\Sigma^+$, let\fxnote{this is called regular cover complexity
of $L$, $\mathrm{REGcc}(L)$ in~\cite{Hetzl19Cover}}
\[
g_\Sigma(L)=
\min\bigl\{|P(G)|:
G\text{ is an acyclic regular grammar over }\Sigma
\text{ and }L\subseteq L(G)\bigr\}.
\]
\end{definition}

\begin{definition}[Positional-cover number]
    For a finite language $L \subseteq \Sigma^+$, define
    \[
    \pc(L)=
    \min\bigl\{|Y|:Y\subseteq\Fact(L)
    \text{ and }Y\text{ positionally covers }L\bigr\}.
    \]
    Here, $Y$ positionally covers $L$ if every position of
    every word in $L$ belongs to an occurrence of a factor
    in $Y$. Unlike $\rank(L)$, this definition does
    not require the factors to concatenate to form each word
    of $L$, and these factors can overlap with one another.
\end{definition}

\subsection{Binary Encoding}\label{sec.binary_encoding}

Let\fxnote{the things explained in this paragraph should be made more formal} $L\subseteq\Sigma^+$ be a finite language and let $ B$ be an arbitrary natural number. Let $p = \lceil \log_2 |\Sigma|\rceil.$, For every $s \in \Sigma$ choose a distinct binary string $x_s$ of length $p$ and let $\mathrm{id}(s)$ be the string obtained by replacing every $0$ with $00$ and every $1$ with $10$ in $x_s$. Thus every identifier has length $2p$, all identifiers are distinct, and none contains 11. Every identifier also ends in $0$. put
\[
\mu_s=11\mathrm{id}(s),\qquad c_B(s)=(\mu_s)^B,
\]
and extend $c_B$ to words by concatenation. Define
\[
\widehat L^B=\{c_B(w):w\in L\}.
\]
For a binary string $z$ that occurs as a factor of an element of $\widehat L^B$, define $\pi(z)$ by:
\begin{enumerate}
    \item find all complete markers $\mu_s$ lying wholly inside $z$, from left to right;
    \item replace each complete marker by its source symbol $s$;
    \item collapse every maximal run of consecutive equal symbols to one copy.
\end{enumerate}
Every complete marker in an encoded input word is aligned with an actual marker. Since all markers have the same length and distinct identifiers, the complete markers contained in $z$ are uniquely determined by $z$ itself thus $\pi$ is well defined.
\begin{lemma}
    Assume that no word of $L$ contains two equal consecutive symbols. If $z$ is a factor of $c(w)$ for some $w \in L$, then
    \[
    \pi(z) \in \Fact(L) \cup \{\eps\}.
    \]
\end{lemma}
\begin{proof}
    Because $z$ is one contiguous substring of $c(w)$, the complete markers lying wholly inside $z$ must be consecutive markers in this sequence. The substring $z$ may cut through a marker at its left or right end, but such incomplete markers are ignored by the definition of $\pi$ and the complete markers in $z$ represent repetitions of consecutive symbols of $w$. Collapsing those repetitions therefore produces a contiguous factor of $w$, or the empty word if no complete marker occurs. Hence
    \[
    \pi(z) \in \Fact(L) \cup \{\eps\}.
    \]
\end{proof}

\begin{lemma}\label{lem.REGcc_encoded}
    Assume that no word of $L$ contains two equal consecutive symbols. Then:
    \begin{enumerate}[label=(\roman*)]
    \item $g_{\{0,1\}}(\widehat L^B)\leq g_\Sigma(L)$.
    \item If $g_{\{0,1\}}(\widehat L^B)\leq B$, then
    \[
    \pc(L)\leq g_{\{0,1\}}(\widehat L^B).
    \]
\end{enumerate}
\end{lemma}

\begin{proof}
    For (i), take an acyclic regular grammar over $\Sigma$ covering $L$ and replace every production's terminal label $w$ by $c_B(w)$. This preserves acyclicity and the number of productions, and the resulting grammar covers $\widehat L^B$.
    \\
    For (ii), let $G$ be an acyclic regular grammar over $\{0,1\}$ covering $\widehat L_B$ with $|P(G)| \leq B$.
    Fix $w=s_1\cdots s_\ell\in L$ and a derivation of $c_B(w)$ in $G$. Let
    $z_1,\ldots,z_r$ be the terminal labels contributed by the productions
    used in that derivation, so that
    \[
    c_B(w)=z_1\cdots z_r.
    \]
    Acyclicity implies that no production is used more than once in a
    derivation, hence
    \[
    r\leq |P(G)|\leq B.
    \]
    There are at most $B-1$ boundaries between consecutive labels. Each
    source symbol $s_i$ is represented by $B$ consecutive copies of its
    marker. Since at most $B-1$ boundaries can cut markers, at least one
    copy of $\mu_{s_i}$ lies wholly inside a single label $z_j$.
    
    Let $Y_G$ consist of all nonempty strings $\pi(z)$ for terminal labels
    of productions used in the selected derivations of words in $\widehat L_B$.
    Then $|Y_G|\leq |P(G)|$, and the preceding observation shows that every
    position of every source word is covered by a factor in $Y_G$. Therefore
    \[
    \pc(L)\leq |Y_G|\leq |P(G)|
    =g_{\{0,1\}}(\widehat L^B).
    \]
    Let $H=(V,E)$ be a graph of maximum degree at most three. Let $L_H$ be the
    language from Section~\ref{sec.reduction_unbalph}. For each edge
    $e=(u,v)$, oriented according to the proper four-coloring used there, the corresponding four words are
    \[
    S_e=\{um_ev,\;um_eb_e,\;a_em_ev,\;a_em_eb_e\},
    \qquad
    L_H=\bigcup_{e\in E}S_e.
    \]
\end{proof}

\begin{lemma}\label{lem.LH_pc}
    For the language $L_H$,
    \[
    \pc(L_H)=\rank(L_H)=3|E|+\tau(H).
    \]
\end{lemma}

\begin{proof}
    In Lemma~\ref{lem.rank_LH} we have proved that $\mathrm{rank(L_H) = 3|E|+\tau(H)}.$  Since $\pc(L_H) \leq \rank(L_H)$ therefore it is easy to see that \[
    \pc(L_H)\leq 3|E|+\tau(H).
    \]
    It remains to prove the following lower bound
    \[
    \pc(L_H)\geq 3|E|+\tau(H).
    \]
    Let $Y\subseteq\Fact(L_H)$ be a positional cover, and let $S=Y\cap V$. For each edge $e=(u,v)$ consider
    \[
    Y_e=\{y\in Y:y\text{ contains at least one of }a_e,m_e,b_e\}.
    \]
    Every nonempty factor of a word in $L_H$ that contains no private letter is
    a singleton vertex letter. Moreover, a factor containing a private letter
    from edge $e$ cannot contain a private letter from another edge. Hence these
    sets partition $Y$ and
    \[
    |Y|=|S|+\sum_{e\in E}|Y_e|.
    \]
    For an edge $e=(u,v)$. Its four words are $um_ev$, $um_eb_e$, $a_em_ev$, and $a_em_eb_e$.
    To cover the final $b_e$ position of $um_eb_e$, the set $Y_e$ must contain a factor
    from
    \[
    \mathcal B=\{b_e,mb_e,um_eb_e\}.
    \]
    To cover the initial $a_e$ position of $a_em_ev$, it must contain a factor from
    \[
    \mathcal A=\{a_e,a_em_e,a_em_ev\},
    \]
    and to cover the middle $m_e$ position of $um_ev$, it must contain a factor from
    \[
    \mathcal M=\{m_e,um_e,m_ev,um_ev\}.
    \]
    These three families are pairwise disjoint. Consequently, $|Y_e|\geq 3$.
    \\
    \\
    We next show that $|Y_e|\geq 4$ whenever neither $u$ nor $v$ belongs to
    $S$. Suppose, for a contradiction, that $|Y_e|=3$. The preceding argument
    and the disjointness of $\mathcal A,\mathcal B,\mathcal M$ imply that $Y_e$
    contains exactly one factor from each family. Since the
    singleton factors $u$ and $v$ are absent, covering the initial $u$-position
    and the final $v$-position of $umv$ forces the factor chosen from $\mathcal M$ to be $um_ev$.
    
    Now the initial $u$-position of $um_eb_e$ can only be covered, without the singleton $u$, by $um_e$ or $um_eb_e$. The factor $um_ev$ does not occur in $um_eb_e$,
    and the chosen factor from $\mathcal A$ contains $a_e$, so the factor chosen
    from $\mathcal B$ must be $um_eb_e$.
    
    Similarly, the final $v$-position of $a_em_ev$
    can only be covered, without the singleton $v$, by $m_ev$ or $a_em_ev$. Neither
    $um_ev$ nor $um_eb_e$ occurs in $a_em_ev$, so the factor chosen from $\mathcal A$ must
    be $a_em_ev$. But none of $umv_e$, $um_eb_e$, and $a_em_ev$ can cover the initial
    $a_e$-position of $a_emb_e$, which requires one of $a_e$, $a_em_e$, or $a_em_eb_e$. This is a
    contradiction. Thus $|Y_e|\geq 4$ when neither endpoint belongs to $S$.
    \\
    \\
    Let $U(S)$ be the set of edges with no endpoint in $S$. then
    \[
    |Y|=|S|+\sum_{e\in E}|Y_e|
    \geq |S|+3|E|+|U(S)|.
    \]
    The set $S$ together with one chosen endpoint of each edge in $U(S)$ is a
    vertex cover, so $\tau(H)\leq |S|+|U(S)|$. It follows that
    \[
    |Y|\geq 3|E|+\tau(H),
    \]
    and hence $\pc(L_H)\geq 3|E|+\tau(H)$.
\end{proof}

\subsection{Reduction}

\begin{theorem}
    Regular Cover is NP-Complete for a fixed alphabet of size 2.
\end{theorem}
\begin{proof}
Membership in NP is clear.
For NP-hardness let $(H,k)$ be an input for the Vertex Cover Problem where $H=(V, E)$ is a graph 
 with $n$ vertices of maximum degree 3 and with m edges.
Let $H'=(V', E')$ be the disjoint union of 7 copies of $H$ then $H$ has a vertex cover of size
 less than or equal to $k$ if and only if $H'$ has a vertex cover of size less than or equal to 
 $7k$.
Then
\[
|E(H')|=7m,\qquad \tau(H')=7\tau(H).
\]
The graph language $L_{H'}$ has an alphabet of size $21m + 7n$. Set
\[
N=21m+7k+6.
\]
Let $\widehat L= \widehat{L_{H'}}^N = \{c_N(w) : w \in L_{H'} \}$ as defined 
 in Section~\ref{sec.binary_encoding}.
Now we will prove that $H$ has a vertex cover of size less than $k$ if and only if there is
 an acyclic regular grammar $G$ such that $\widehat L \subseteq L(G)$ and $|P(G)| \leq N$.

For the left-to-right direction suppose that $H$ has a vertex cover of size less than $k$.
Then $H'$ has a vertex cover of size less than $7k$ then by Lemma~\ref{lem.rank_LH}
\[
\rank(L_{H'})=3|E(H')|+\tau(H') = 21m+7\tau(H)\leq 21m+7k.
\]
By Lemma~\ref{lem.lchained_grammar} since there is 6 different levels in
 $I(\lambda) \cup O(\lambda)$, there is an acyclic regular grammar over the alphabet
 $\{0, 1\}$ covering $L_{H'}$ with at most
\[
\rank(L_{H'})+6\leq 21m+7k+6=N
\]
productions.
Replacing each terminal label $w$ by $c(w)$ gives an acyclic regular grammar over the
 alphabet grammar covering $\widehat L$ with at most $N$ productions.
 
For the right-to-left direction suppose that there exists an acyclic regular grammar $G$
 such that $\widehat{L} \subseteq L(G)$ and $|P(G)| \leq N$.
By Lemma~\ref{lem.REGcc_encoded} we have
\[
\pc(L_{H'})\leq |P(G)|\leq N.
\]
Therefore by Lemma~\ref{lem.LH_pc}
\[
\pc(L_{H'}) = \rank(L_{H'}) = 21m+7\tau(H) \leq 21m+7k+6.
\]
It follows that
\[
\tau(H)\leq k+\frac67<k+1.
\]
\end{proof}

%

\bibliographystyle{plain}
\bibliography{bibliography}

\end{document}